\documentclass[
reprint,
amsmath,amssymb,
aps,
prb,
]{revtex4-2}

\usepackage{graphicx}
\usepackage{dcolumn}
\usepackage{bm}

\begin{document}
\title{Interface-induced band bending and charge separation in all-organic ZnPc/F$_x$ZnPc heterostructures} 
	
\author{Stephanie Amos}
\affiliation{Department of Physics and Astronomy, University of Kansas, Lawrence, KS 66045, USA}
\author{Neno Fuller}
\affiliation{Department of Physics and Astronomy, University of Kansas, Lawrence, KS 66045, USA}
\author{Wai-Lun Chan}
\affiliation{Department of Physics and Astronomy, University of Kansas, Lawrence, KS 66045, USA}
\author{Hartwin Peelaers}
\affiliation{Department of Physics and Astronomy, University of Kansas, Lawrence, KS 66045, USA}
\email{peelaers@ku.edu}

\date{\today}

\begin{abstract}
Organic semiconductors are attractive building blocks for electronic devices due to their low cost and flexibility. Furthermore, heterostructures with type-II band alignments can efficiently separate photogenerated charges via a charge transfer and separation process. 

Here, we use density functional theory (DFT) to investigate model interfaces formed by zinc phthalocyanine (ZnPc) and its fluorinated derivatives (F$_8$ZnPc and F$_{16}$ZnPc). We demonstrate that these interfaces not only exhibit a type-II band offset, but also band bending. The band bending causes both the LUMO and HOMO states to localize away from the interface. Therefore, the band bending creates a strong driving force for charge separation. We used ultraviolet photoemission spectroscopy (UPS) to experimentally confirm this predicted band bending. The wavefunction envelopes of vertically-stacked molecules resemble particle-in-a-box states, but this shape is lost when the molecules are staggered.

These results elucidate how interface-induced band bending facilitates charge separation in all-organic heterostructures and suggest a design pathway toward improved performance in organic photovoltaic devices.

\end{abstract}

\maketitle

\section{\label{sec:level1}Introduction}

Organic semiconductors are an attractive low-cost, flexible building block for (opto)electronic device applications~\cite{Li2018, Deng2023}. For organic photovoltaics it is important to separate photogenerated excitons through charge transfer (CT) and charge separation (CS) processes. A well-known CT process in inorganic 2D materials is the creation of heterostructures with type-II band alignments.
All-organic heterostructures combine the benefits of improved CT from type-II band alignments, while keeping the low cost and flexibility from their organic nature. 

We focus on heterostructures based on phthalocyanines (Pcs), as these form a tunable and computationally efficient model system. Pc molecules are composed of four isoindole units linked by eight nitrogen atoms. These Pcs have sixteen hydrogen atoms around the perimeter of the molecule and two hydrogen atoms at the center of the molecule. The geometric and electronic properties of these molecules can be tuned by replacing the two center hydrogen atoms with a single metal atom, such as Li, Na, K, or Zn, and thereby creating a metal phthalocyanine (MPc)~\cite{Berkowitz1979, Nuleg2018}. Such MPcs allow promising device applications including microelectronics, chemical sensors, single crystal field-effect transistors, information storage, photocatalysis, and biosensors~\cite{Li2018, Deng2023}. 2D zinc phthalocyanine (ZnPc) materials demonstrate excellent electronic properties for use in solar cell technology~\cite{Jiang2017}. Greater tuning can be achieved through fluorination, in which a number of the perimeter hydrogen atoms are replaced with fluorine atoms~\cite{Oison2007, Schwarze2016, Schlettwein2001, Kafle2021}. Large ($\geq$ 8mm) samples of ZnPc crystals can be created with a low-cost, green, solvothermal process which adds to the appeal of these molecules~\cite{Li2018}.

\begin{figure}[t]
	\includegraphics[width=0.95\columnwidth]{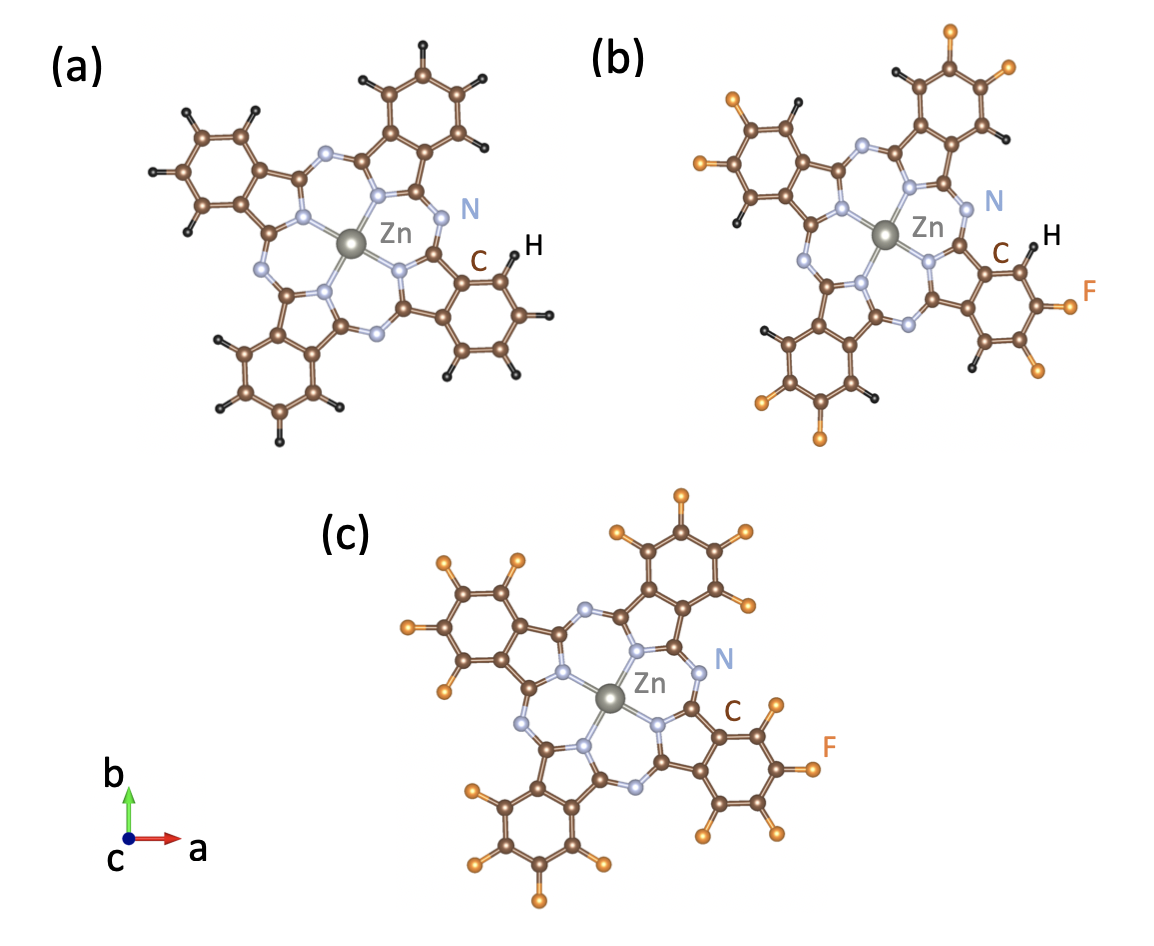}
	\caption{\label{fig:TopView} Top view of (a) ZnPc,  (b) F$_{8}$ZnPc, and (c) F$_{16}$ZnPc.}
\end{figure}

\begin{figure*}[t]
\includegraphics[width=1.6\columnwidth]{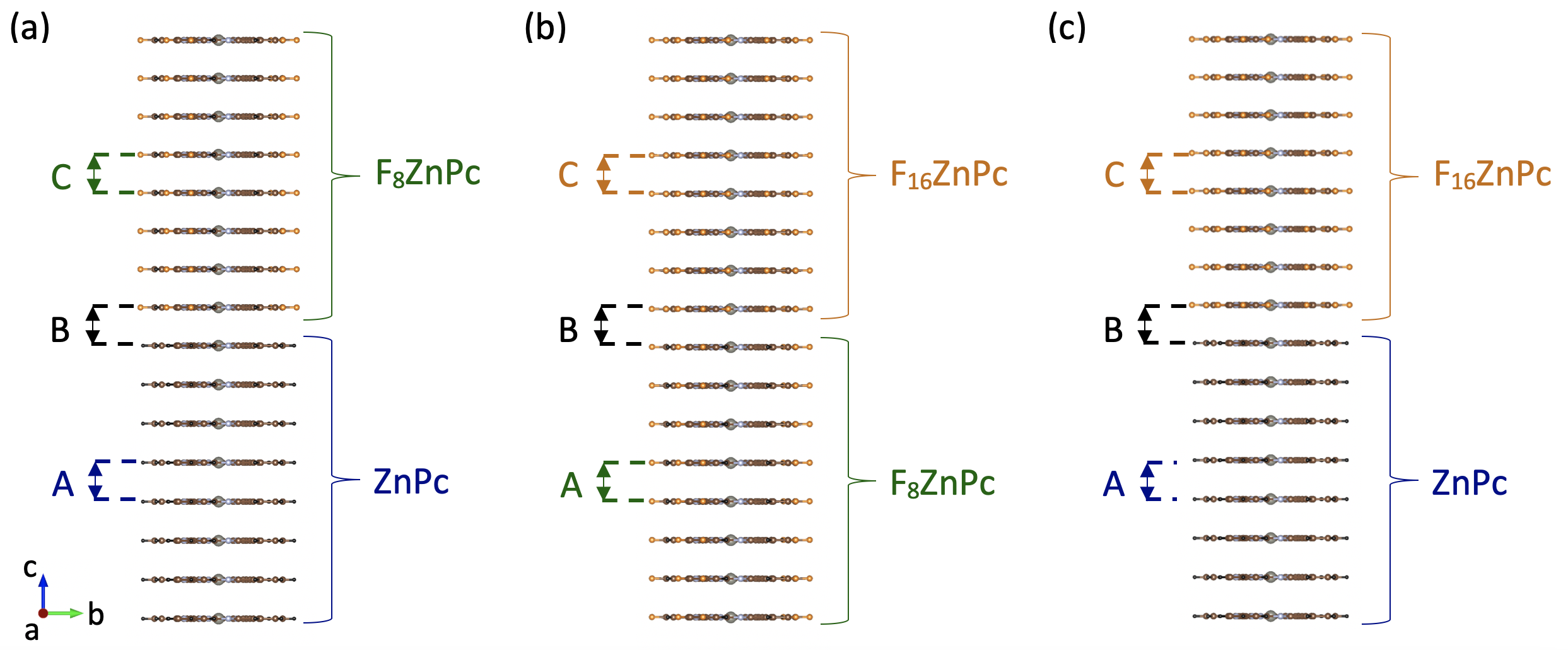}
\caption{\label{fig:structure} Side views of (a) ZnPc and F$_{8}$ZnPc, (b) F$_{8}$ZnPc and F$_{16}$ZnPc, and (c) ZnPc and F$_{16}$ZnPc stacking configurations. (Distance between molecules at positions A, B, and C are provided in Table~\ref{tab:Spacing}.)}
\end{figure*}

Several heterostructures with ZnPc have been analyzed in an effort to improve CT and subsequent exciton charge separation (CS) which is critical for current flow in efficient solar cells. Some of these heterostructures include ZnPc on MoS$_{2}$~\cite{Kafle2019, Liu2018, Ulman2021},  ZnPcF$_{16}$ on p-6p OTFTs~\cite{Sun2017}, and ZnPc with F$_{4}$-TCNQ~\cite{Gao2001, Gao2002}. Understanding the driving mechanisms for CT and CS in such heterostructures is important for designing higher efficiency solar cells. One theory proposes that exciton fission, in which a hot exciton decays into multiple cold excitons, will lead to increased efficiency due to the increase in the number of excitons~\cite{Berkelbach2013}. Contrary to this theory, a recent study identified that delocalized hot excitons are more advantageous for current generation than multiple localized cold excitons~\cite{Kafle2018}. 
Information relating to CT excitons, exciton localization, and CS can be deduced through a variety of electronic properties of various heterostructures. As an example, CT and CS mechanisms were explored for ZnPc on MoS$_{2}$ through the band bending of the valence band maximum (VBM) and conduction band minimum (CBM) at the material interface~\cite{Liu2018, Kafle2019}. Exciton charges in ZnPc on monolayer MoS$_{2}$ are able to delocalize, while exciton charges from ZnPc on bulk MoS$_{2}$ will recombine because of large interfacial band bending~\cite{Kafle2019}. 
Heterostructures composed of ZnPc with F$_{8}$ZnPc have CT excitons, where electrons move from the ZnPc molecules to the F$_{8}$ZnPc molecules~\cite{Kafle2021}. 

Precise details of the charge separation mechanisms for many ZnPc-based heterostructures are still missing. These heterostructures can serve as model system for the highly efficient non-fullerene acceptors (NFA) organic photovoltaics, where band bending was proposed as charge separation mechanism~\cite{Karuthedath2021}.
In contrast to these systems, ZnPc-based heterostructures are computationally more efficient due to the smaller sizes of the molecules (57 atoms/molecule). Here, we use density functional theory (DFT) to simulate large cells with 16 F$_x$ZnPc molecules, allowing us to investigate the band bending and charge separation. We show that the band bending is such that both electron (LUMO) and hole (HOMO) wavefunctions will localize away from the interface, leading to charge separation. In model vertical stackings, the LUMO and HOMO resemble particle-in-a-box wavefunctions. However, in more realistic staggered stackings, which reduce the heterostructure energy, this particle-in-the-box character is lost. Our predicted band bending agrees well with experimental findings. These quantitative results elucidate the causes of the charge separation process, which can be used as a design pathway to improve the performance of all-organic photovoltaics.

\section{Methods}
We use density functional theory (DFT) with projector augmented wave (PAW) potentials~\cite{Blochl1994} as implemented in the Vienna Ab-initio Simulation Package (VASP)~\cite{Kresse1993,Kresse1996}. We use the meta-GGA R2SCAN functional~\cite{Furness2020} in combination with Grimme D4 van der Waals corrections~\cite{Caldeweyher2017}. The plane wave expansion cutoff was set at 400 eV and we used the $\Gamma$ point to sample the Brillouin zone. Atomic positions in the individual molecules were relaxed until the forces on the atoms were less than 5 meV/Å. The initial intermolecular spacing was identified by fitting the energy as a function of intermolecular spacing and using the obtained minimum energy. These preliminary results were refined through relaxation of the structure until forces (including intermolecular forces) were below 20 meV/Å. We used 1x10$^{-7}$ eV as energy convergence criterion. 

Projected density-of-states (DOS) and charge densities were obtained from VASP and rendered visually using Pymatgen~\cite{Jain2011,Ong2013}.
Structures are visualized using the \textsc{VESTA} code~\cite{Momma2011}.

Per molecule HOMO (HOMO$_i$) and LUMO (LUMO$_i$) energies are obtained from the projected DOS~\cite{Pham2025,Giustino2005}, by solving
\begin{equation}\label{eq:VBMfit}
	\int_{HOMO_i}^{E_F} D(\epsilon,z)d\epsilon = \Delta_H \int_{-\infty}^{E_F} D(\epsilon,z)d\epsilon.
\end{equation}
\begin{equation}\label{eq:CBMfit}
	\int_{LUMO_i}^{E_F} D(\epsilon,z)d\epsilon = \Delta_L \int_{-\infty}^{E_F} D(\epsilon,z)d\epsilon,
\end{equation}

where D($\epsilon$,z) is the projected density for each molecule, $\epsilon$ the energies, $z$ the molecule position along the vertical c-axis (as shown in Fig.~\ref{fig:structure}), and E$_{F}$ the Fermi energy. $\Delta_H$ and $\Delta_L$ are obtained from a fit to the same equations, but replacing HOMO$_i$ and LUMO$_i$ with the global HOMO and LUMO energies.

\section{Results}

We study three different stacking configurations for ZnPc and fluorinated ZnPc. The first stacking configuration consists of ZnPc with F$_{8}$ZnPc, the second consists of ZnPc with F$_{16}$ZnPc, and the third consists of F$_{8}$ZnPc with F$_{16}$ZnPc. These stacking configurations are shown in Fig.~\ref{fig:structure}. For each structure, the less fluorinated molecules form the base of the stack, while the more highly fluorinated molecules form the top of the stack. 

\begin{table}[h]
	\caption{Interlayer distance in \AA~ at positions A, B, and C as indicated in Fig.~\ref{fig:structure} and Fig.~\ref{fig:StagStruct}} 
	\centering 
	\begin{tabular}{c c c c c} 
		\hline\hline \\
		Position & & A & B & C
		\\ [0.5ex]
		\hline\hline 
		F$_{8}$ZnPc on ZnPc &   & 3.69 & 3.61 & 3.61 \\
		\hline 
		F$_{16}$ZnPc on F$_{8}$ZnPc &   & 3.60 & 3.57 & 3.57  \\
		\hline 
		F$_{16}$ZnPc on ZnPc &   & 3.67 & 3.64 & 3.60 \\
		\hline 
		Staggered &   &  &  &  \\
		F$_{8}$ZnPc on ZnPc &   & 3.34 & 3.33 & 3.33 \\
		\hline 
	\end{tabular}
	\label{tab:Spacing}
\end{table}

\subsection{ZnPc with F$_{8}$ZnPc and F$_{8}$ZnPc with F$_{16}$ZnPc}
\begin{figure}[tbh]
	\includegraphics[width=0.95\columnwidth]{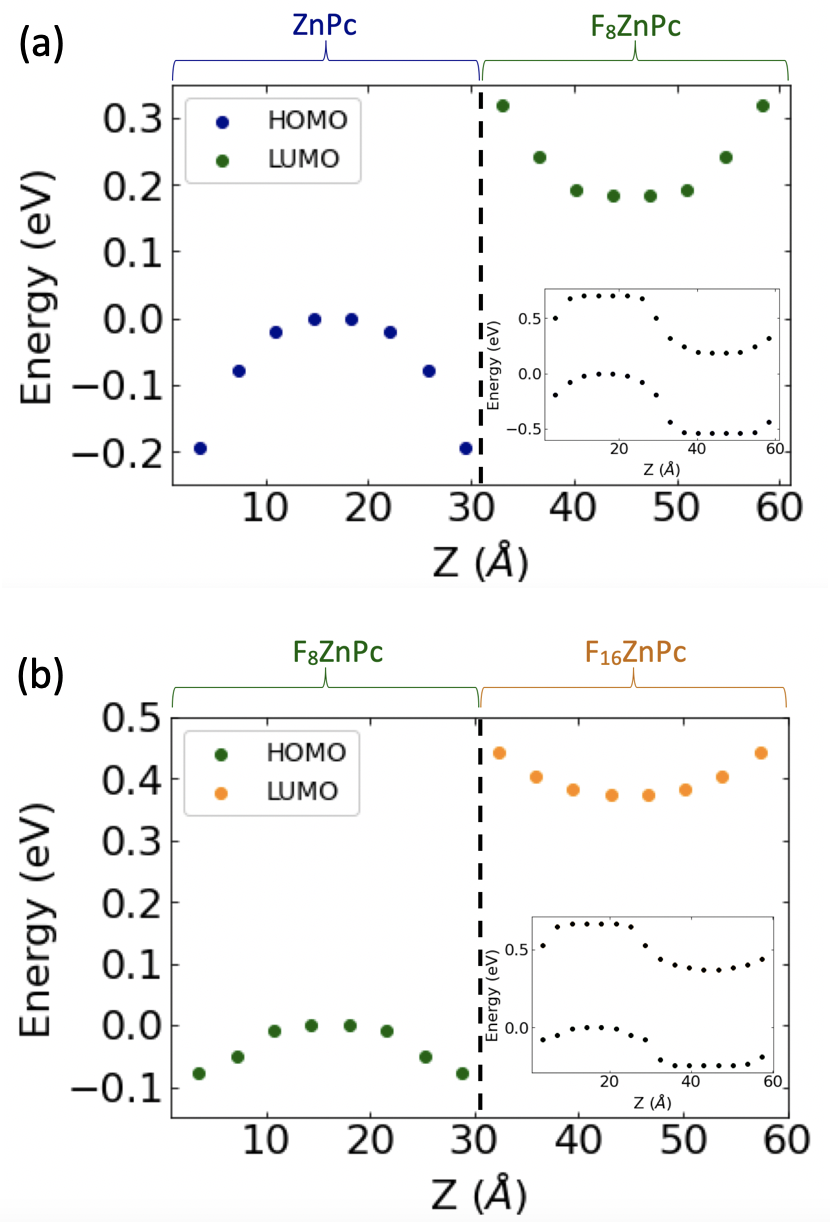}
	\caption{\label{fig:EngBands} Band bending at (a) the ZnPc/F$_{8}$ZnPc interface and (b) the F$_{8}$ZnPc/F$_{16}$ZnPc interface. The insets show the full view of HOMO and LUMO levels.}
\end{figure}

F$_{8}$ZnPc on ZnPc structure and F$_{16}$ZnPc on F$_{8}$ZnPc structure (Fig.~\ref{fig:structure}(a) and (b)) have very similar properties.
The band bending of the HOMO and LUMO levels are shown in Fig.~\ref{fig:EngBands} (a) for the F$_{8}$ZnPc on ZnPc structure and in Fig.~\ref{fig:EngBands} (b) for the F$_{16}$ZnPc on F$_{8}$ZnPc structure. The band gaps and band widths for these materials are listed in Table~\ref{tab:BandInfo}.
The overall HOMO level is located on the base layer molecules, while the overall LUMO level is located on the top layer molecules for both heterostructures. This alignment agrees with previous research on CT excitons~\cite{Manaka2003, Kafle2021}, although the band gaps are smaller than in experiment. 

Both materials exhibit band bending with the HOMO level occurring at interior molecules in the base layers of the structure (less fluorinated molecules) and the LUMO occurring at interior molecules in the top layers of the structure (more fluorinated molecules). The F$_{16}$ZnPc on F$_{8}$ZnPc structure has a larger band gap and smaller band widths compared to the F$_{8}$ZnPc on ZnPc structure. 

\begin{table}[h]
\caption{Band gaps and band widths for F$_{8}$ZnPc on ZnPc and F$_{16}$ZnPc on F$_{8}$ZnPc in eV}
\centering 
\begin{tabular}{c c c} 
\hline\hline 
  &  F$_{8}$ZnPc on ZnPc &  \\ 
\hline 
Band Gap & HOMO Band Width & LUMO Band Width \\
 & (ZnPc) & (F$_{8}$ZnPc) \\
0.18 & 0.19 & 0.14 \\
\hline\hline  
  &  F$_{16}$ZnPc on F$_{8}$ZnPc &  
\\ [0.5ex]
\hline 
Band Gap & HOMO Band Width & LUMO Band Width \\
 & (F$_{8}$ZnPc) & (F$_{16}$ZnPc) \\
0.37 & 0.08 & 0.07 \\
\hline\hline\hline 
  & Staggered F$_{8}$ZnPc on ZnPc &  \\ 
\hline 
Band Gap & HOMO Band Width & LUMO Band Width \\
 & (ZnPc) & (F$_{8}$ZnPc) \\
0.68 & 0.21 & 0.19 \\
\hline 
\end{tabular}
\label{tab:BandInfo}
\end{table}

\begin{figure}[h]
	\includegraphics[width=0.95\columnwidth]{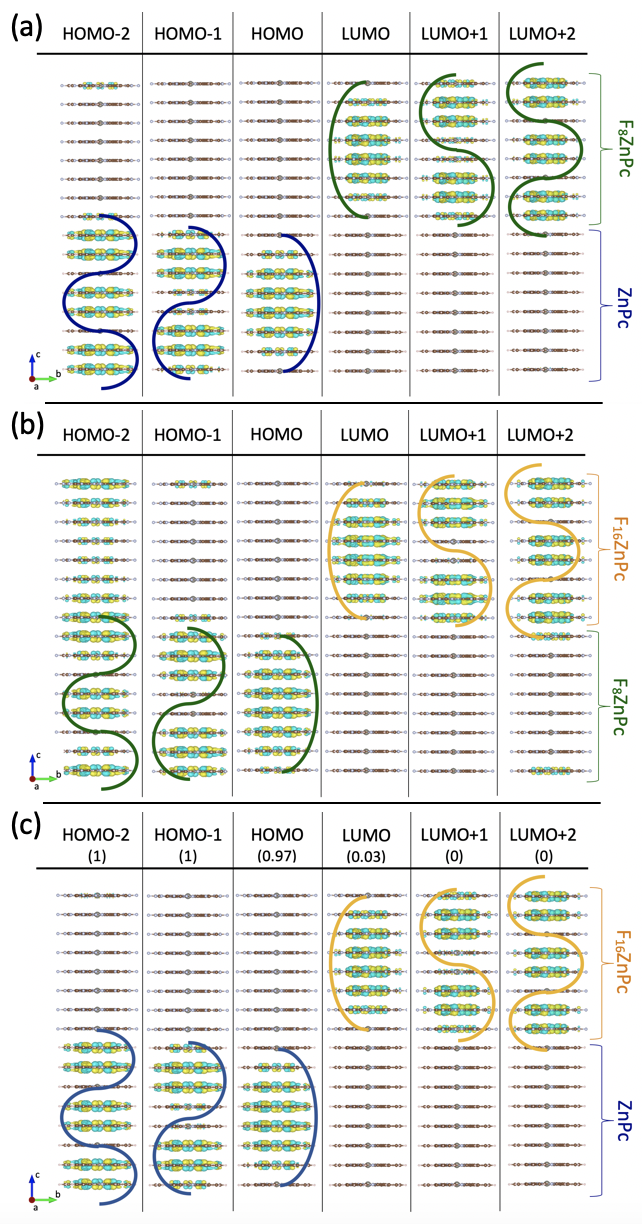}
	\caption{ Wavefunctions for the three highest occupied and lowest unoccupied states at $\Gamma$ for (a) the ZnPc/F$_{8}$ZnPc interface, (b) the F$_{8}$ZnPc/F$_{16}$ZnPc interface, and (c) the ZnPc/F$_{16}$ZnPc interface. The corresponding interface structures are shown in Fig.~\ref{fig:structure}. The solid lines are a schematic guide to the eye, tracing out the envelope function of the wavefunctions.
\label{fig:wavefunctions}	
}
\end{figure}

We can obtain more insights by visualizing the wavefunctions corresponding to the three highest occupied states and the three lowest unoccupied states at the $\Gamma$ point, as shown in Fig.~\ref{fig:wavefunctions} (a) and (b). All occupied states are localized on the base layer molecules (ZnPc in the ZnPc/F$_{8}$ZnPc structure and F$_{8}$ZnPc in the F$_{8}$ZnPc/F$_{16}$ZnPc structure), and all unoccupied states are localized on top layer molecules. This is consistent with the band bending picture of Fig.~\ref{fig:EngBands}. Interestingly, the overall shape of the F$_{8}$ZnPc on ZnPc wavefunctions (schematically depicted with solid lines) closely resemble the solutions of a particle-in-a-box, for holes (HOMOs) and electrons (LUMOs). Lower lying states (holes) and higher lying states (electrons) show an increasing number of nodes, as expected from excited states. This is consistent with the energy profile shown in Fig.~\ref{fig:EngBands}, where holes are confined in an energy well in ZnPc and electrons in an energy well in F$_8$ZnPc for the F$_8$ZnPc stacking on ZnPc. Similar wavefunctions are observed in the F$_{16}$ZnPc on F$_{8}$ZnPc structure [Fig.~\ref{fig:wavefunctions}(b)] with the exception of some more significant wavefunctions localizing in the top layer F$_{16}$ZnPc molecules for the HOMO-2 case and very small wavefunction localization into the base layer F$_{8}$ZnPc molecules for the LUMO+2 case.

\begin{figure}[tbh]
	\includegraphics[width=0.95\columnwidth]{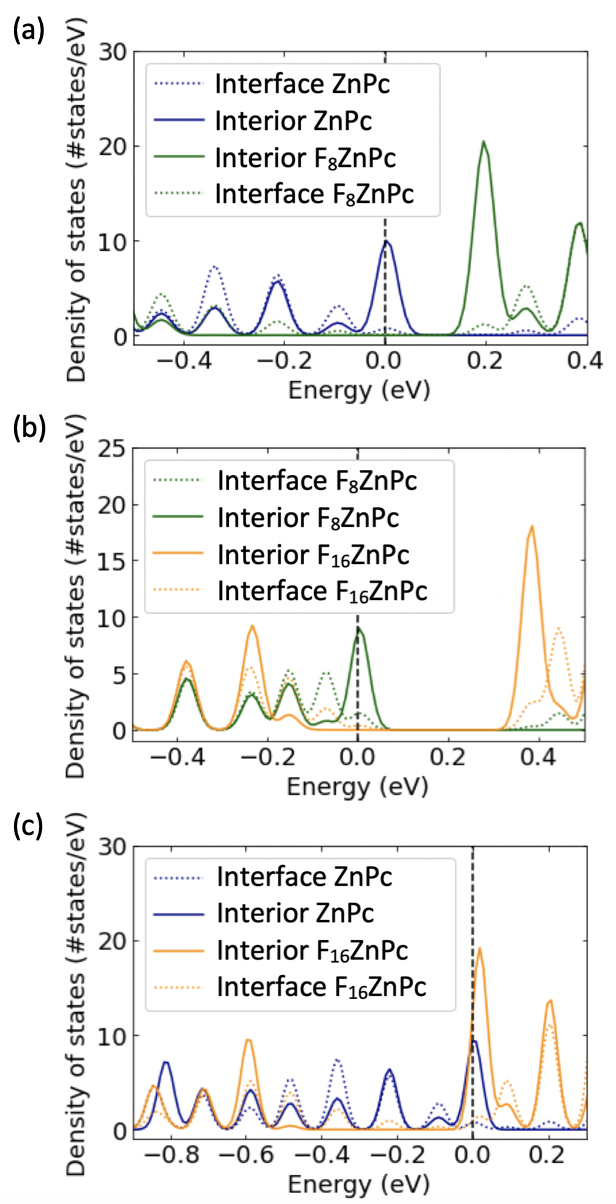}
	\caption{\label{fig:DOS} The projected density of states for interface molecules (dashed lines) and interior molecules (solid lines) in the vertical stacks consisting of (a) 8 ZnPc and 8 F$_8$ZnPc molecules, (b) 8 F$_8$ZnPc and 8 F$_{16}$ZnPc molecules, and (c) 8 ZnPc and 8 F$_{16}$ZnPc molecules.}
\end{figure}

Fig.~\ref{fig:EngBands} also shows that the interior base layer molecules HOMO level is at a higher energy relative to the interface base layer molecules, while the interior top layer molecules LUMO levels are at lower energy states relative to the interface top layer molecules. 
To further confirm this, we calculated the density of states (DOS), projected on either interface or interior molecules (see Fig.~\ref{fig:DOS}(a) and (b)). The zero of energy is set at the HOMO level. For both structures the DOS corresponding to the two topmost HOMO levels is indeed located on the interior base layer molecules, while the DOS corresponding to the two lowest LUMO levels is located on the interior top layer molecules. 

\subsection{Staggered molecules}
\begin{figure}[t]
	\includegraphics[width=0.95\columnwidth]{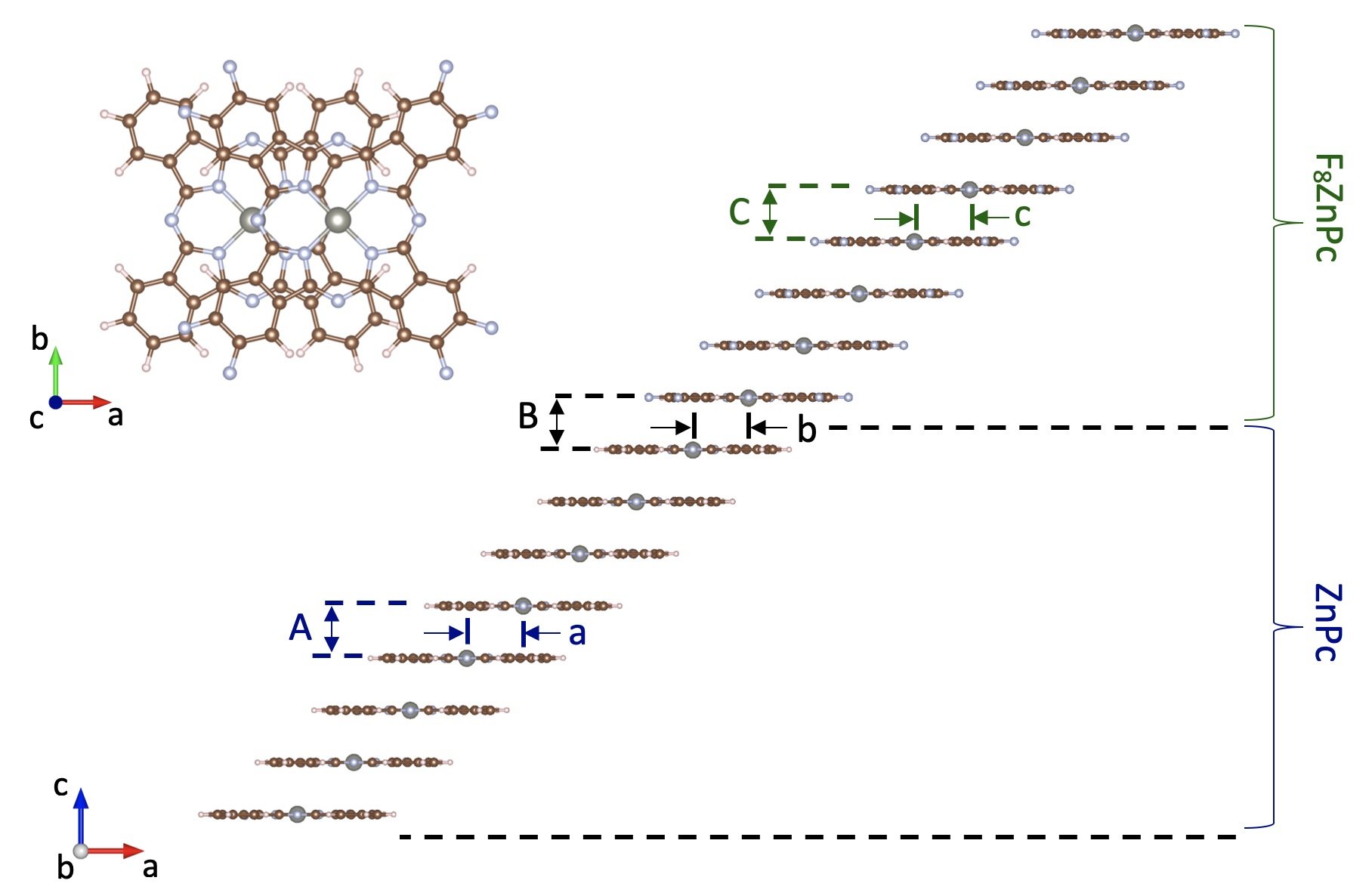}
	\caption{\label{fig:StagStruct} Staggered structure with top view of two molecules showing an offset along the \textbf{a}-axis. Interlayer spacing: A=3.34 \AA, B=3.33 \AA, C=3.33 \AA; Offset: a=3.62 \AA, b=3.63 \AA, c=3.54 \AA.}
\end{figure}

Perfectly stacked molecules form a nice model system, but in reality ZnPc and F$_{8}$ZnPc will stack in a staggered fashion, as such staggering lowers the energy. To identify the effects of this staggering we consider a staggered structure for F$_8$ZnPc on ZnPc. Our calculations indicate that the free energy is lowered by almost 4 eV in the staggered stacking as shown in Fig.~\ref{fig:StagStruct} compared to the model vertical stacking used before. We will therefore elucidate the effects of this more realistic stacking. 

We first quantify the optimal staggering and intermolecular distances. For our initial structure, we offset each molecule by a distance equal to the intermolecular distance creating a 45\textdegree ~slipping angle~\cite{Nuleg2018}. We then follow the methodology described in Section II to further optimize the structures. The resulting distances and offsets are listed in Fig.~\ref{fig:StagStruct}. 

\begin{figure}[tbh]
	\includegraphics[width=0.95\columnwidth]{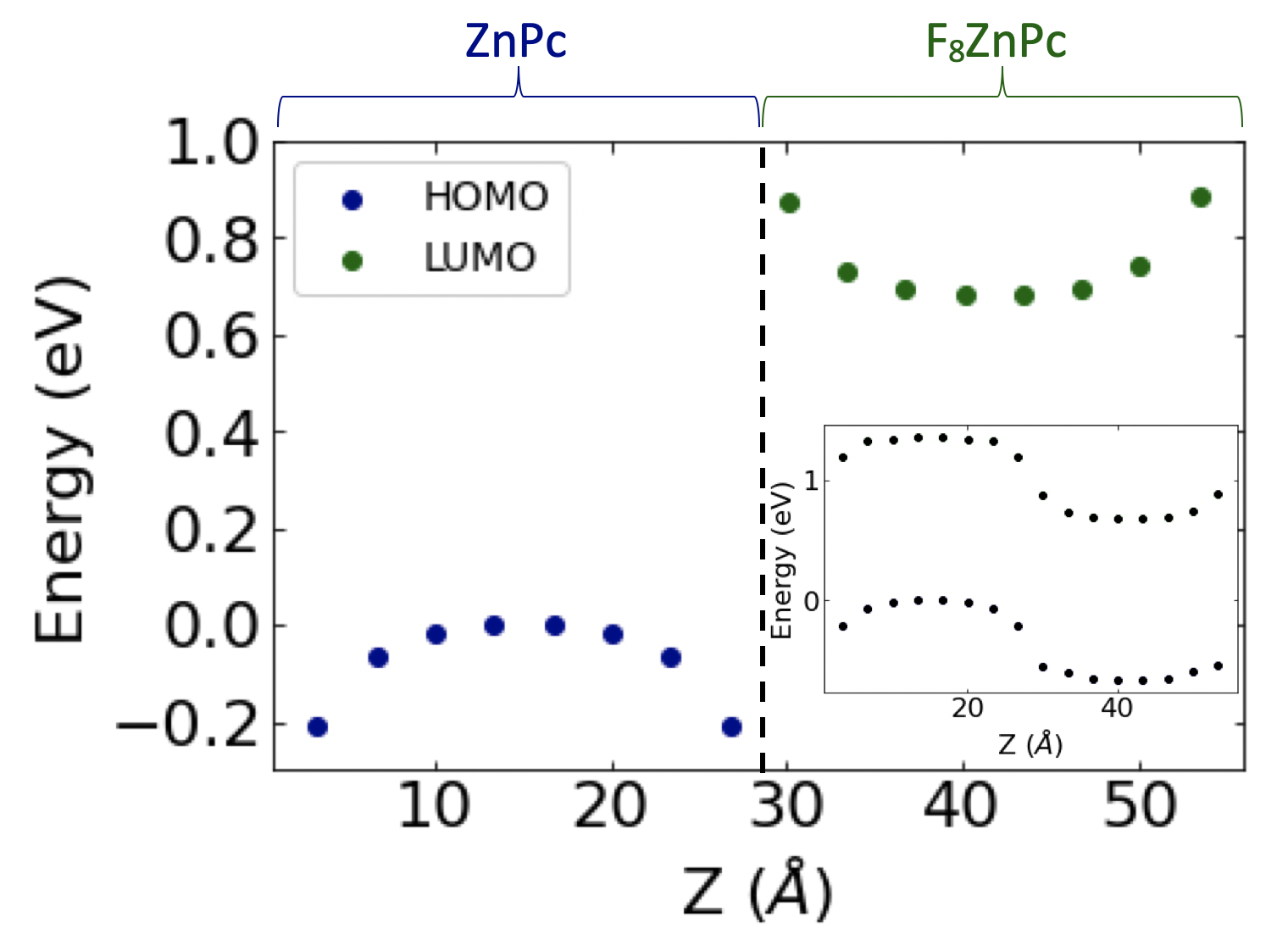}
	\caption{\label{fig:StagEngBandsZnF8} Band bending for the staggered structure at the ZnPc/F$_{8}$ZnPc interface. The inset shows the full view of HOMO and LUMO levels.}
\end{figure}

Next, we calculate the band bending for the staggered structure (see Fig.~\ref{fig:StagEngBandsZnF8}). The band bending at the interface for the staggered configuration is similar to the bending seen in Fig.~\ref{fig:EngBands} (a) for the model vertical stacked system. However, for the staggered structure, the band gap is increased by 0.5eV (see Table~\ref{tab:BandInfo}), which brings the calculated band gap closer to experimental results, further indicating that staggered structures are more realistic. Note that the magnitude of the band bending also increases for staggered structures.

\begin{figure}[tbh]
	\includegraphics[width=\columnwidth]{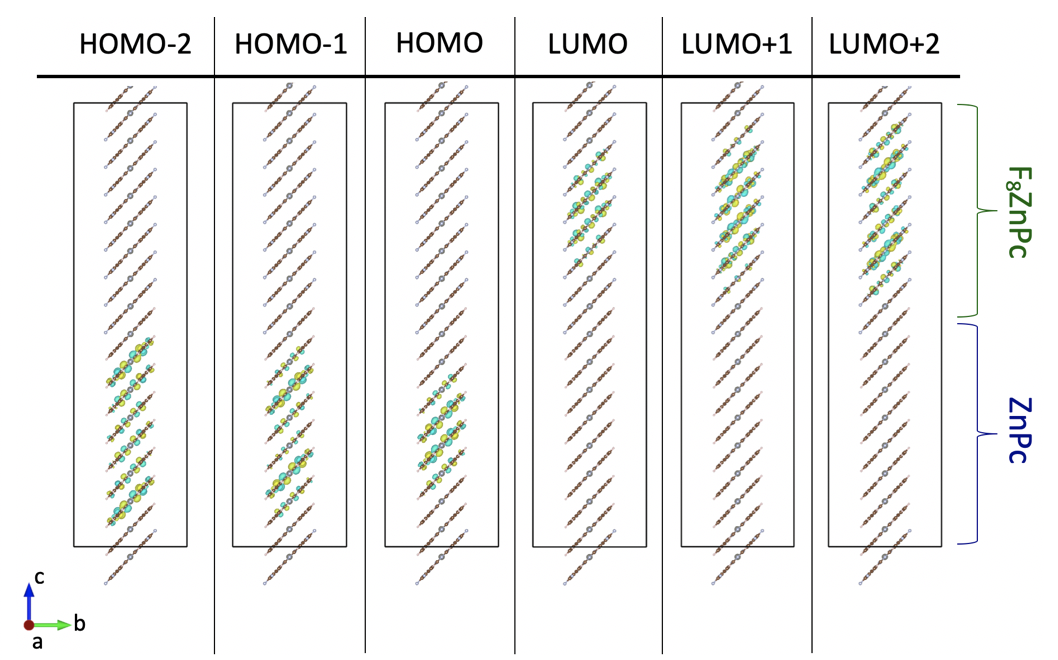}
	\caption{\label{fig:StagWF} Wavefunctions for the three highest occupied and lowest unoccupied states at $\Gamma$ for the staggered stacking configuration.}
\end{figure}

\begin{figure}[tbh]
	\includegraphics[width=0.95\columnwidth]{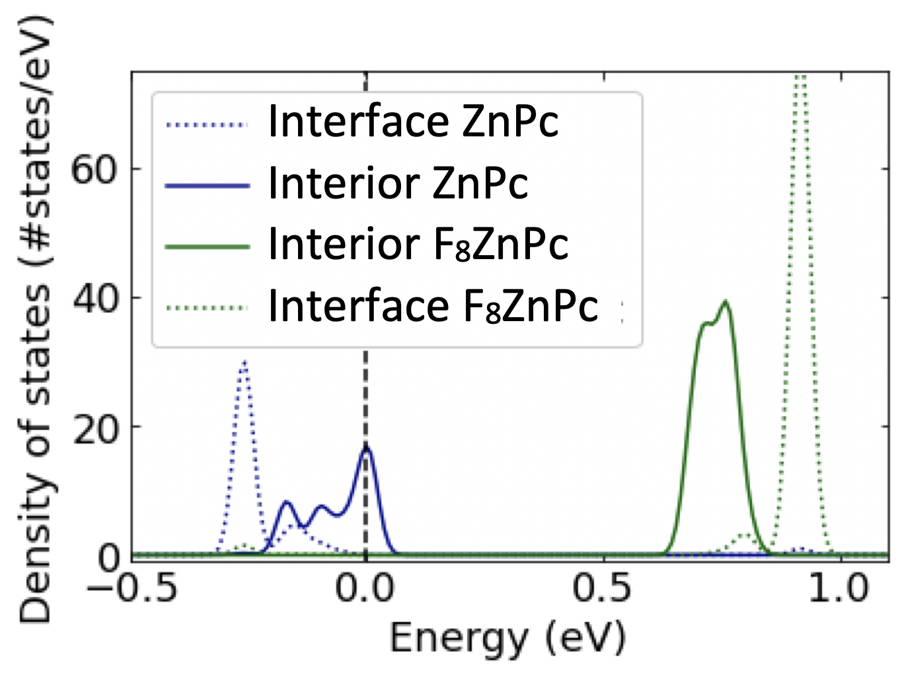}
	\caption{\label{fig:StagDOS} The density of states for a staggered stack consisting of 8 ZnPc and 8 F$_8$ZnPc molecules projected on interface molecules (dashed lines) or interior molecules (solid lines).}
\end{figure}

As expected from the band-bending picture, the HOMO and LUMO states are located on the interior molecules, as illustrated by the projected DOS shown in Fig.~\ref{fig:StagDOS}. The wavefunction plots (Fig.~\ref{fig:StagWF}) for the HOMO and LUMO wavefunctions also show the localization on the interior molecules.

Wavefunctions shown in Fig.~\ref{fig:StagWF} are considered distinct wavefunctions if their energy differences are larger than 0.01 eV. The resulting HOMO and LUMO orbital energy differences are in the range of 0.01 eV to 0.04 eV for the first six HOMO and LUMO levels. 
The staggering results in changes in the envelope functions of the wavefunctions so that these no longer resemble particle-in-the-box states. However, the wavefunctions are still more strongly localized on interior molecules, as expected from the increased band bending.

These results, for both vertically stacked and staggered ZnPc and F$_8$ZnPc, show that upon light excitation, it is energetically favorable for holes to reside in ZnPc and for electrons to reside in F$_8$ZnPc. Moreover, there is also a driving force for either charge carrier to move away from the interface (as that lowers the energy of both holes and electrons). The formation of charge-transfer excitons is therefore likely, with charges localized away from the interface. In turn, this will make it easier to separate charges in photovoltaic devices, thereby increasing their performance.  

\subsection{ZnPc with F$_{16}$ZnPc}
Interestingly, the HOMO to LUMO gap vanishes in the F$_{16}$ZnPc on ZnPc structure with partially occupied orbitals at the interface. Similar to the F$_{8}$ZnPc on ZnPc wavefunctions, the overall shape of the F$_{16}$ZnPc on ZnPc wavefunctions [solid lines in Fig.~\ref{fig:wavefunctions}(c)] closely resembles the solutions of a particle-in-a-box, for holes (HOMOs) and electrons (LUMOs). Note that the terms HOMO and LUMO are only approximate due to the energy overlap (the numbers in parenthesis indicate the electron occupation of these states), as is also visible in the projected DOS [Fig.~\ref{fig:DOS}(c)]. Since the near-LUMO level projected on F$_{16}$ZnPc is partially filled and the near-HOMO level projected on ZnPc is partially empty, some ground-state charge transfer~\cite{Manaka2003} will occur. Similar transfer has been observed in TTF/TCNQ heterostructures~\cite{Alves2008,Moayedpour2023,Kattel2017}.
Since the staggered stacking for the similar F$_{8}$ZnPc on ZnPc material increased its bandgap by 0.5 eV compared to the vertical stacking, it is possible that a similar effect may cause a bandgap to emerge for a staggered configuration of F$_{16}$ZnPc on ZnPc.

\section{Experimental results}
To support the calculations, we also performed experimental measurements on F$_8$ZnPc/ZnPc and F$_4$ZnPc/ZnPc heterostructures. While we did not explicitly calculate F$_4$ZnPc/ZnPc heterojunctions, we expect to observe similar band bending. Bilayer F$_x$ZnPc on ZnPc films are grown on a graphite substrate in ultrahigh vacuum environment. The energy of the ZnPc’s HOMO and F$_x$ZnPc’s HOMO was measured by ultraviolet photoemission spectroscopy (UPS) for different thicknesses of the top F$_x$ZnPc layer. Because the UPS method is surface sensitive and it typically only probes the top molecular layer, it allows us to measure the HOMO energy as a function of distance away from the interface. Detailed experimental procedures can be found in Ref.~\cite{Kafle2021,Fuller2025}. 

\begin{figure}[tbh]
	\includegraphics[width=0.95\columnwidth]{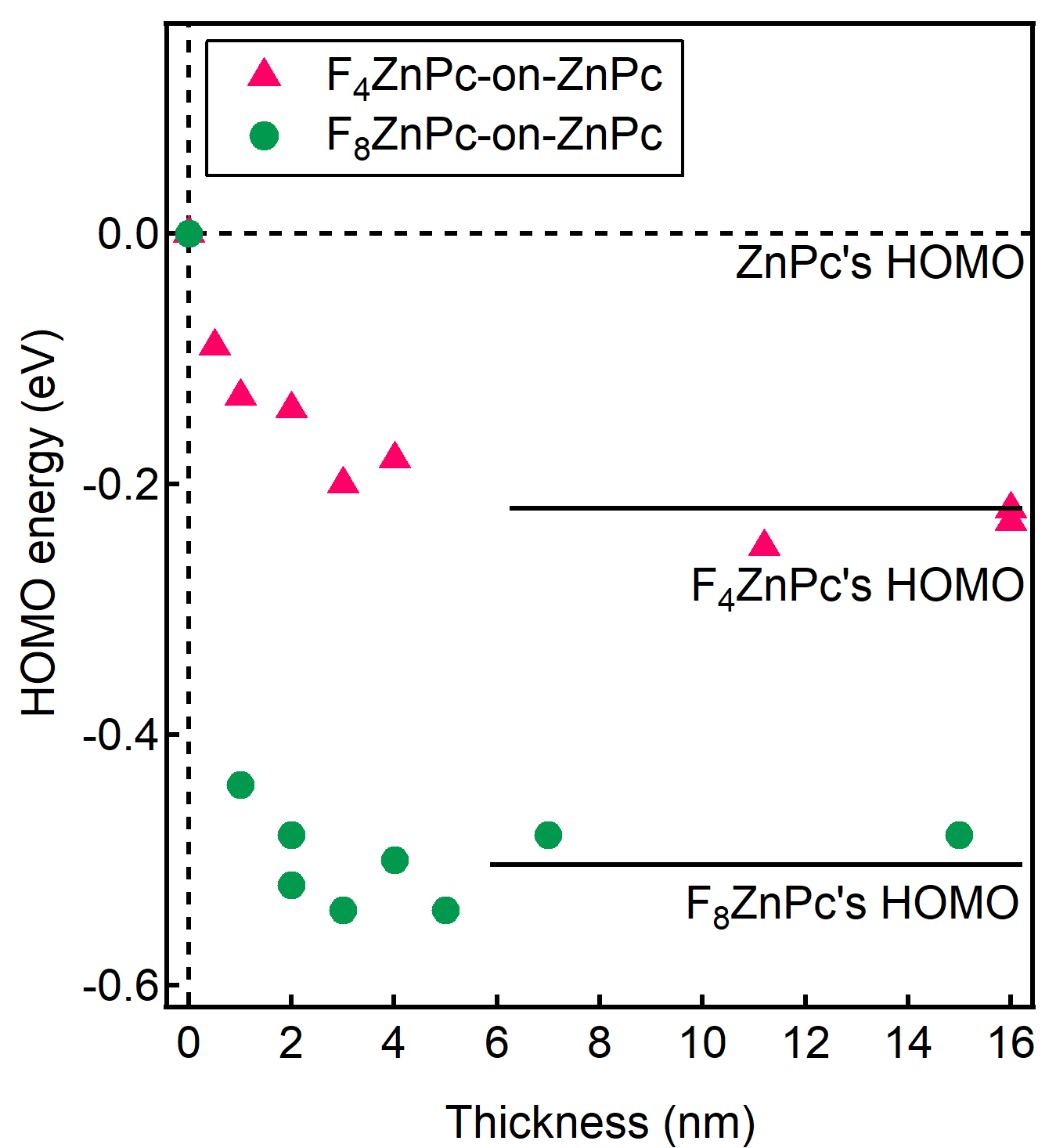}
	\caption{\label{fig:exp} The HOMO energy of F$_x$ZnPc as a function of the F$_x$ZnPc’s thickness. The F$_x$ZnPc layer is grown on top of a 10-nm ZnPc layer. The HOMO energy of ZnPc is set at 0 eV.}
\end{figure}

Fig.~\ref{fig:exp} shows the measured HOMO energy versus the F$_x$ZnPc’s thickness. This thickness essentially represents the distance away from the ZnPc layer. In this plot, the position of ZnPc’s HOMO is set at 0 eV. After the growth of an ultrathin (~0.5 nm) layer of F$_x$ZnPc, we first observe a sharp drop in the HOMO energy, which can be interpreted as the band offset between ZnPc and F$_x$ZnPc. Then, the HOMO level continues to shift downward by an amount of ~ 0.1 eV for both F$_8$ZnPc/ZnPc and F$_4$ZnPc/ZnPc, which agrees very well with the amount of band bending determined by the DFT calculations [see Table~\ref{tab:BandInfo} and Fig.~\ref{fig:EngBands}(a)].

\section{Conclusion}
We used DFT calculations to show that heterostructures of ZnPc, F$_8$ZnPc, and F$_{16}$ZnPc not only exhibit a band offset (except vertically stacked ZnPc/F$_{16}$ZnPc heterostructures), but also band bending at the interface. This band bending is so that both HOMO and LUMO wavefunctions will localize away from the interface. We find that F$_8$ZnPc and ZnPc both prefer to stack in a staggered fashion, where their centers are offset with respect to each other. If the molecules are stacked vertically, the envelope functions of the wavefunctions resemble particle-in-a-box states, but this envelope-function shape is lost when staggering the molecules.
UPS measurements on ZnPc/F$_8$ZnPc and ZnPc/F$_4$ZnPc confirm the predicted band bending.
Our results show that upon photoexcitation, a driving force exists for electrons and holes to localize away from the interface. Such charge separation should increase the performance of organic photovoltaic devices. 

\section*{Acknowledgement}
This work is supported by the U.S. Department of Energy, Office of Science, Office of Basic Energy Sciences, Chemical Sciences, Geosciences, and Biosciences Division under Award DE-SC0024525. This research used resources of the National Energy Research Scientific Computing Center, a DOE Office of Science User Facility supported by the Office of Science of the U.S. Department of Energy under Contract DE-AC02-05CH11231 using NERSC Award BES-ERCAP0033524. Additional computational resources were provided by the University of Kansas Center for Research Computing (CRC), including the BigJay Cluster resource funded through NSF Grant MRI-2117449.

\bibliography{BandBendingZnPc}

\end{document}